\documentclass[oneside,letter,12pt,pdflatex]{article}
\usepackage[utf8]{inputenc}
\usepackage{float}

\usepackage{setspace}

\usepackage[top=1in, bottom=1in, left=1in, right=1in]{geometry}

\usepackage[section]{placeins}

\usepackage[usenames,dvipsnames]{xcolor}
\usepackage{apacite}
\definecolor{winered}{rgb}{0.5,0,0}
\usepackage[colorlinks=true, linkcolor=black, citecolor=winered, urlcolor=black, bookmarks=true]{hyperref}
\hypersetup{urlcolor=blue}

\usepackage[round]{natbib}

\usepackage{amsmath}
\usepackage{amsthm}
\usepackage{amsfonts}
\usepackage{thmtools}
\usepackage{amssymb}

\usepackage{graphicx}
\graphicspath{ {figures/} }
\usepackage{caption}
\usepackage{longtable}

\usepackage{booktabs}
\usepackage{tabulary}
\usepackage{lscape}
\usepackage[flushleft]{threeparttable}
\usepackage[flushleft]{threeparttablex}
\usepackage{xcolor}

\newcommand{\beginsupplement}{%
        \setcounter{table}{0}
        \renewcommand{\thetable}{S\arabic{table}}%
        \setcounter{figure}{0}
        \renewcommand{\thefigure}{S\arabic{figure}}%
     }

\title{Program Targeting with Machine Learning and Mobile Phone Data: Evidence from an Anti-Poverty Intervention in Afghanistan{\footnote{\small We thank Seungmin Lee, Maria Camila Ayala and Thomas Escande for excellent research assistance. This work was supported by DARPA and NIWC under contract N66001-15-C-4066, the NSF under grant IIS-1942702, and by the World Bank's Knowledge for Change Program. The U.S. Government is authorized to reproduce and distribute reprints for Governmental purposes not withstanding any copyright notation thereon. The views, opinions, and/or findings expressed are those of the authors and should not be interpreted as representing the official views or policies of the Department of Defense, the U.S. Government, the World Bank and its affiliated organizations, or those of the Executive Directors of the World Bank or the governments they represent.}}}

\author{Emily L. Aiken\footnote{\small School of Information, University of California, Berkeley} 
\and Guadalupe Bedoya\footnote{\small Development Impact Evaluation Department, World Bank} 
\and Joshua E. Blumenstock\footnotemark[2] 
\and Aidan Coville\footnotemark[3]}

\date{\small This version: \today\\ First version: April 10, 2020}

\begin{document}

\begin{spacing}{1}

\maketitle

\begin{abstract}

Can mobile phone data improve program targeting? By combining rich survey data from a ``big push'' anti-poverty program in Afghanistan with detailed mobile phone logs from program beneficiaries, we study the extent to which machine learning methods can accurately differentiate ultra-poor households eligible for program benefits from ineligible households. We show that machine learning methods leveraging mobile phone data can identify ultra-poor households nearly as accurately as survey-based measures of consumption and wealth; and that combining survey-based measures with mobile phone data produces classifications more accurate than those based on a single data source. 
\end{abstract}

\vspace{1cm}
\vspace{2cm}

\end{spacing}

\newpage


\section{Introduction}
\label{sec:introduction}

Each year, hundreds of billions of dollars are spent on targeted social protection programs. The importance of these programs increased dramatically in the past 18 months: In 2020, global extreme poverty increased for the first time in two decades, and most countries expanded their social protection programs, with more than 1.1 billion new recipients receiving government-led social assistance payments \citep{gentilini_social_2020}. 

Determining who should be eligible for program benefits --- \textit{targeting} --- is a central challenge in the design of these programs \citep{hannaolken,lindert2020sourcebook}. In high-income countries, targeting frequently relies on tax records or other administrative data on income. In low- and middle-income countries (LMICs), where a large fraction of the workforce is informal, programs often require primary data collection. The difficulty and cost of collecting data, and the variable quality of what gets collected, can introduce significant errors in the targeting process \citep{Deaton2016,jerven, wb_forthcoming}. These issues are exacerbated in fragile and conflict-affected countries, where two thirds of the world's poor are expected to reside by 2030 \citep{corral2020}. 

This paper evaluates the extent to which non-traditional administrative data, processed with machine learning, can be used for program targeting. Specifically, we match call detail records (CDR) from a large mobile phone operator in Afghanistan to household survey data from the Afghan government's Targeting the Ultra-Poor (TUP) anti-poverty program. Eligibility for the TUP program was determined through a \textit{hybrid targeting method}, combining  a community wealth ranking (CWR) and a short follow-up survey. We then assess the accuracy of three counterfactual targeting approaches at identifying the actual beneficiaries of the TUP program: (i) our \textit{CDR-based method}, which applies machine learning to data from the mobile phone company; (ii) an \textit{asset-based wealth index}, which uses asset ownership to approximate poverty; and (iii) \textit{consumption}, a common benchmark for measuring poverty in LMICs.

Our analysis produces three main results. First, by comparing errors of inclusion and exclusion using the program's hybrid method as a benchmark, we find that the CDR-based method is nearly as accurate as the commonly-employed asset and consumption-based methods for identifying the phone-owning ultra-poor households. Second, we find that methods combining CDR data with measures of assets and consumption are  more accurate than methods using any single data source. Third, we find that when non-phone-owning households are included in the analysis, the CDR-based method remains accurate if non-phone-owning households are classified as ultra-poor; however, targeting performance is quite poor if households without phones are ineligible for benefits. After presenting these main results, we compile data from several existing targeting programs to give an indication of the substantial reduction in marginal costs associated with CDR-based targeting.

These results connect two distinct strands of prior work. The first is a literature on program targeting, which studies the effectiveness of different mechanisms for identifying program beneficiaries. In LMICs, research has focused on the performance of proxy means tests (PMTs) \citep{grosh1995proxy,filmer2001,brown2018}, community-based targeting strategies (CBTs) \citep{alatas2012,Fortinetal2016}, and related approaches \citep{banerjee2007,karlan2013,Premandetal2020}. A meta-analysis by \citet{coady2004}, which includes 8 PMTs and 14 community-based programs, finds little difference in targeting accuracy between the two methods --- but notes that targeting is regressive in a quarter of programs reviewed. In addition to issues with targeting accuracy, the current methods available for poverty targeting in LMICs are time- and resource-intensive, and  may be infeasible in fragile or conflict-affected areas or in contexts where social interaction is limited, such as during a pandemic. 


The second body of work explores the extent to which non-traditional sources of data, in conjunction with machine learning, might help address data gaps in LMICs  \citep[e.g.,][]{blumenstock2016,burke2021}. Much of this work focuses on estimating the geographic distribution of poverty at fine spatial granularity, using data from satellites \citep{jean_combining_2016,engstrom2017poverty}, mobile phones \citep{blumenstock2015,hernandez}, social media \citep{fatehkia2020mapping,ermon2019}, or some combination of these data sources \citep{steele2017,pokhriyal,chi_micro-estimates_2020}.
Most relevant to our current analysis, two prior papers investigate whether the mobile phone use can approximate the wealth of individual mobile subscribers. \cite{blumenstock2015} show that CDR data are predictive of an individual-level asset-based wealth index among a nationally representative sample of 856 Rwandan mobile phone owners ($r$ = 0.68).  \cite{blumenstock2018} finds similar results with a sample of 1,234 male heads of households in the Kabul and Parwan districts of Afghanistan. While these results show that phone data can be used to \textit{predict} poverty levels, they do not evaluate whether those poverty estimates are of sufficient quality for real-world policy applications. 

Our paper connects these two literatures by rigorously assessing the extent to which phone-based estimates of poverty can help with program targeting \citep{blumenstock2020}.
We believe the analysis will be especially relevant to the increasing number of interventions that rely on mobile money to distribute cash payments \citep{gentilini_social_2020}, and the growing number of contexts where mobile phone data are being made available for humanitarian purposes \citep{milusheva2021challenges}. For example, in just the past few years, mobile money was used to make cash transfer payments in countries including Bangladesh \citep{mm_bangladesh}, Ghana \citep{mm_ghana}, Liberia \citep{mm_liberia}, and Malawi \citep{mm_malawi}. Mobile phone \textit{data} has been used to guide cash transfers in Colombia \citep{gentilini_social_2020}, the Democratic Republic of the Congo \citep{gentilini2021cash}, Pakistan \citep{gentilini_social_2020}, and Togo \citep{aiken_machine_2021}.%
\footnote{The anti-poverty program implemented in Togo and described by \citet{aiken_machine_2021} was based on the methods developed and evaluated in this paper. Due to the time-sensitive nature of the COVID-19 response described in \citet{aiken_machine_2021}, the two academic articles are in circulation concurrently.}

The context of our empirical analysis -- identifying ultra-poor households in Afghanistan -- is a particularly challenging environment for data collection and program targeting, as 62\% of the households classified as not \textit{ultra}-poor still fall below the national poverty line. In such environments, when traditional options for targeting are not feasible, these methods may provide a viable alternative for identifying households with the greatest need. Given the policy relevance of these results, we conclude our analysis by discussing important ethical and logistical considerations that may influence how CDR methods are used to support targeting efforts in practice.

\section{Data and Methods}
\label{sec:methods}

\subsection{Household Survey Data}
\label{section:groundtruth}
Our empirical analysis relies on survey data collected as part of the Targeting the Ultra-Poor (TUP) program implemented by the government of Afghanistan with support from the World Bank. The TUP program was a ``big push," providing multi-faceted benefits to 7,500 ultra-poor households in six provinces of Afghanistan between 2015 and 2018 \citep{bedoya2019}. Our analysis uses data from the baseline and targeting surveys for an impact evaluation of the TUP program conducted in 80 of the poorest villages of Balkh province between November 2015 and April 2016 (N=2,582).\footnote{Our analysis restricts to 2,814 households for whom consumption and asset data are non-missing.}

\paragraph{Ultra-Poor Designation} Eligibility for the TUP program was determined based on based on geographic criteria,\footnote{The poorest villages were identified by the availability of veterinary services, financial institutions, and social services, and being relatively accessible \citep{bedoya2019}.}
followed by a two-step process including a community wealth ranking (CWR) and a follow-up in-person survey. CWRs were conducted separately in each village, coordinated by a local NGO and village leaders, in collaboration with the government team. 
The CWR was followed by an in-person survey to determine whether nominated households met a set of qualifying criteria, coordinated by the NGO and government representatives, and based on a measure of multiple deprivation. 

For a household to be designated as \textit{ultra-poor}, and therefore eligible for program benefits, it had to be considered extreme-poor in the CWR (43\% of households), and also meet at least three of six criteria:%
\begin{enumerate}
\item Financially dependent on women's domestic work or begging
\item Owns less than 800 square meters of land or living in a cave
\item Primary woman under 50 years old
\item No adult men income earners
\item School-age children working for pay
\item No productive assets
\end{enumerate}

Ultimately, 11\% of the households classified as extreme-poor in the community wealth ranking step --- 6\% of the total population in the study villages --- were classified as ultra-poor and eligible for TUP benefits. Of the 2,852 households surveyed for the TUP project, 1,173 (41\%) were designated as ultra-poor, and 1,679 (59\%) were non-ultra-poor.%
\footnote{Surveys were conducted following the CWR and eligibility verification. The response rate for ultra-poor households was 96\%. Approximately 20 households in each of the study villages were randomly drawn (excluding TUP-eligible households), to provide a representative benchmark for the TUP sample.} 

\paragraph{Consumption} The consumption module of the TUP survey contains information on food consumption for the week prior to the interview and non-food expenditures for the year prior to the interview. These are used to construct monthly \textit{per capita} consumption values, as detailed in \cite{bedoya2019}. Based on these data, we measure the logarithm of \textit{per capita} monthly consumption, using the same approach that the Afghan government used to determine the national poverty line.  

\paragraph{Asset Index} We construct an asset-based wealth index to assess the relative socioeconomic status of surveyed households, calculated as the first principal component of variation in household asset ownership for sixteen items in Table \ref{appendix:firstcomponent}. The principal component analysis is calculated across the 2,814 households with complete asset data, after standardizing each asset variable to zero mean and unit variance. This wealth index explains 25.3\% of the variation in asset ownership. Figure \ref{appendix:assetindex} shows the distribution of the underlying asset index components and Table \ref{appendix:firstcomponent} shows the direction of the first principal component. 




\paragraph{Other Variables} The TUP surveys collected several other covariates that we use in subsequent analysis. These include a food security index (composed of variables relating to the skipping and downsizing of meals, separately for adults and children), a financial inclusion index (composed of access to banking and credit, knowledge of banking and credit, and savings), and a psychological well-being index for the primary woman (standardized weighted scores on the Center for Epidemiological Studies Depression scale, the World Values Survey happiness and satisfaction questions, and Cohen's Stress Scale) -- see \citet{bedoya2019}. The survey also collected data from each household on mobile phone ownership. Nearly all (99\%) households with a cell phone provided their phone numbers and consented to the use of their call detail records for this study.   

\paragraph{Sample Representativity} 
\label{section:selectionintosample}
Portions of our analysis are restricted to the 535 households from the TUP survey with phone numbers that match to our CDR (see Section \ref{section:cdr}). Table \ref{selectionintosample} and~\ref{appendix:kdeoutcomes} compare characteristics of these households to the full survey population. 
There are some systematic differences: the 535-household sample is wealthier, which is consistent with households in the subsample being required to own at least one phone.  For instance, while 88\% of non-ultra-poor households in the TUP survey own at least one phone, only 72\% of ultra-poor households own at least one phone. 

\paragraph{Summary Statistics} 
As shown in Table \ref{selectionintosample} and Figure \ref{appendix:correlations}, the  three measures of well-being in our dataset are only weakly correlated with one another: for example, the correlation between the asset index and consumption is 0.37 in the full survey and 0.34 in the matched subsample. Likewise, while the ultra-poor population makes up 27\% of the overall sub-sample, less than half of the ultra-poor fall into the bottom 27\% of the sample by wealth index or consumption.

\paragraph{Sample Weights} Since the TUP survey oversampled the ultra-poor (by a factor of roughly 12), portions of our analysis use sample weights to adjust for population representativity. When sample weights are applied, it is explicitly noted; if not mentioned, no weights are applied. 
After sample weights are applied, the ultra-poor make up 5.98\% of the overall population, and 4.63\% of our matched subsample. 




\subsection{Mobile Phone Metadata} 
\label{section:cdr}

In a follow-up survey conducted in 2018, we requested informed consent from survey respondents to obtain their mobile phone CDR and match them to the survey data collected through the TUP project. CDR contain detailed information on:

\begin{itemize}
    \item \textbf{Calls: } Phone numbers for the caller and receiver, time and duration of the call, and cell tower through which the call was placed
    \item \textbf{Text messages: } Phone numbers for the caller and recipient, time of the message
    \item \textbf{Recharges: } Time and amount of the recharge
\end{itemize}

For participants who consented, we match baseline survey data (collected November 2015 - April 2016) to CDR covering that same period, obtained from one of Afghanistan's main mobile phone operators. For households with multiple phones and a designated household head (N=65), we match to CDR for the phone belonging to the household head. For households where the household head does not have a phone and someone else does (N=17), we match to CDR for one of the households' phones selected at random. In total, for the 535 households in our sample, 629,543 transactions took place in the months of November 2015 to April 2016, broken down into 310,883 calls, 305,756 text messages, and 12,904 recharges.

From these CDR, we compute a set of 797 behavioral indicators that capture aggregate aspects of each individual's mobile phone use \citep{bandicoot}. This set includes indicators relating to an individual's communications (for example, average call duration and percent initiated conversations), their network of contacts (for example, the entropy of their contacts and the balance of interactions per contact), their spatial patterns based on cell tower locations (for example, the number of unique antennas visited and the radius of gyration), and their recharge patterns (including the average amount recharged and the time between recharges). 
The distributions of a sample of these indicators are shown in Figure~\ref{appendix:feature_kdes}.

\subsection{Machine Learning Predictions} 

\paragraph{CDR-based Method} Extending the approach described in \citet{blumenstock2015}, we test the extent to which ultra-poor status can be predicted from CDR. This analysis uses the 535 TUP households who match to CDR to train a supervised machine learning algorithm to predict ultra-poverty status from the mobile phone features. The intuition --- also highlighted in Figure~\ref{appendix:feature_kdes} --- is that ultra-poor individuals use their phones very differently than non-ultra-poor individuals, and machine learning algorithms can use those differences to predict ultra-poor status.

Our main analysis uses a gradient boosting model, which generally out-performs several other common machine learning algorithms for this task (see Table~\ref{appendix:mldetails}). The feature importances for the trained model are shown in Table~\ref{appendix:importances}. To limit the potential for overfitting, probabilistic predictions are generated via 10-fold cross-validation, with folds stratified to preserve class balance.\footnote{While cross validation is a standard evaluation strategy in the machine learning literature, for robustness we present results using a basic single train-test split in Table \ref{onesplit}.} 
Additional details on the machine learning methods are provided in Appendix \ref{appendix:ml}.


\paragraph{Combined Methods} We also evaluate several approaches that use data from multiple sources to predict ultra-poor status. Our main \textit{combined method} trains a logistic regression to classify the ultra-poor and non-ultra-poor households using the predicted ultra-poor probability from the CDR-based method (i.e., the output of the gradient boosting algorithm described above), as well as asset and consumption data collected in the TUP survey. For comparison, we similarly evaluate the performance of methods that combine only two of the available data sources (i.e., assets plus consumption, assets plus CDR, and consumption plus CDR). Predictions for each of the combined methods are pooled over 10-fold cross-validation.

\subsection{Targeting Accuracy Evaluation}
\label{section:evaluation}

\paragraph{Evaluation on Matched Subsample} Our main analysis focuses on the 535 households for which we observe both CDR and survey data, and evaluates whether machine learning methods leveraging CDR data can accurately identify households designated as ultra-poor by the TUP program (using the two-step hybrid approach described in Section~\ref{section:groundtruth}). We compare the performance of the CDR-based method to the performance of methods based on the wealth index, consumption data, and combinations of these data sources.%
\footnote{The CDR-based method uses supervised learning to model the ultra-poverty outcome, whereas the asset- and consumption-based approaches do not. To assess the importance of this difference, we experiment with applying machine learning methods to the asset and consumption data to model the ultra-poverty outcome. In results shown in Table~\ref{appendix:assetbasedpred}, we find that a machine-learned asset predictor provides slight improvements on the standard asset-based wealth index and consumption measures. We continue to use the standard asset and consumption measures as benchmarks in the remainder of the paper, however, as they are the targeting methods most frequently used in practice.}
Each targeting method is evaluated based on classification accuracy, errors of exclusion (ultra-poor households misclassified as non-ultra-poor) and errors of inclusion (non-ultra-poor households misclassified as ultra-poor). 
We focus on the ultra-poor designation as the `ground truth' status of the household, against which other methods are evaluated, since it is the most carefully vetted measure of well-being for this population, and the proxy that the government used to target TUP benefits.

To evaluate the performance of the CDR-based and combined methods, we pool out-of-sample predictions across the ten cross-validation folds, so that every household in our dataset is associated with a CDR-based predicted probability of ultra-poor status that is produced out-of-sample.\footnote{In Table~\ref{onesplit}, we show that results are unchanged when we use a single train-test split, instead of 10-fold cross-validation.} To account for class imbalance, we evaluate model accuracy using a ``quota method", by selecting a cut-off threshold for ultra-poor qualification such that each method identifies the proportion of ultra-poor households in our subsample; this cut-off also balances inclusion and exclusion errors. This quota-based approach reflects a scenario in which a program has a fixed budget constraint; it is  also frequently used in the targeting literature \citep{alatas2012, schnitzer2021targeting}. In our 535-household matched dataset this threshold is 27\%; in other samples (see following subsection), the percentage is different. We evaluate each method for precision (positive predictive value) and recall (sensitivity). To capture the trade-off between inclusion and exclusion errors for varying values of this threshold, we also construct receiver operating characteristic (ROC) curves for each method and consider the area under the curve (AUC) as a measure of targeting quality. For each evaluation metric (precision, recall, and AUC), we bootstrap 1,000 samples from the original dataset to calculate the standard deviation of the mean of the accuracy metric. Each bootstrapped sample is of the same size as the original dataset, drawn with replacement.

\paragraph{Accounting for Households Without Phones} 
Our main results assess the performance of different targeting methods on the sample of 535 households for whom we have both survey data and mobile phone data. 
We also present results that show how performance is affected when the analysis includes TUP households for whom we do not have mobile phone data (typically because they do not have a phone or because they use a different phone network than the one who provided CDR). We provide analysis that targets such households (1) before households with CDR, or (2) after households with CDR (see  Section~\ref{sec:nophones}). These results are evaluated on three different samples:

\begin{enumerate}
    \item \textit{Matched Sample:} The 535 households for whom could match survey responses to CDR.
    
    \item \textit{Balanced Sample:} This sample includes the 535 matched households as well as the 472 households in the TUP survey who report not owning any phone. It excludes households that own a phone on a different phone network than the one who provided CDR. The motivation for this sample is to provide an indication of targeting performance in a regime in which CDR can be used to target all phone-owning households. In addition to applying sample weights from the survey, households that do not own a phone are downweighted so that the balance of phone owners to non-phone-owners (with sample weights applied) is the same as in the baseline survey as a whole (with sample weights applied, 84\% phone owners).

    \item \textit{Full Sample:} All 2,814 households in the TUP baseline survey for which asset and consumption data are available, with sample weights applied.

\end{enumerate}

Note that the quota used to evaluate targeting changes for each sample, based on the number of households that are ultra-poor in the sample. For the matched sample, the targeting quota is 27.29\%; for the balanced sample and full sample the quotas are 5.47\% and 6.02\%, respectively.


\section{Results}
\label{sec:results}

\subsection{Performance of Targeting Methods}
Our first set of results evaluate the extent to which different targeting methods can correctly identify ultra-poor households. This analysis compares the performance of CDR-based targeting methods to asset-based and consumption-based targeting, using the sample of 535 households for which survey data and CDR data are both available. 

\begin{figure}[H]
\centering 
\caption{Predicting ultra-poor status from CDR}.  
\includegraphics[width=1\textwidth]{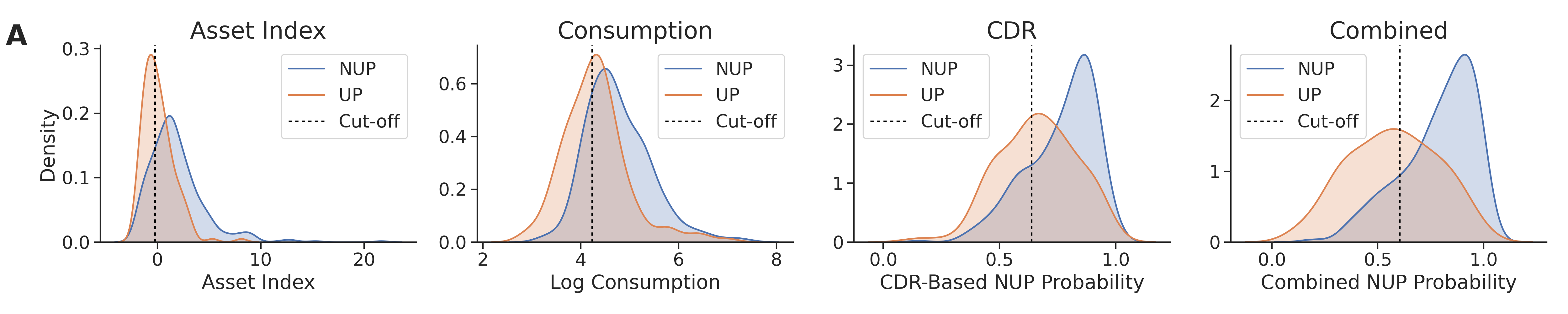}
\includegraphics[width=1\textwidth]{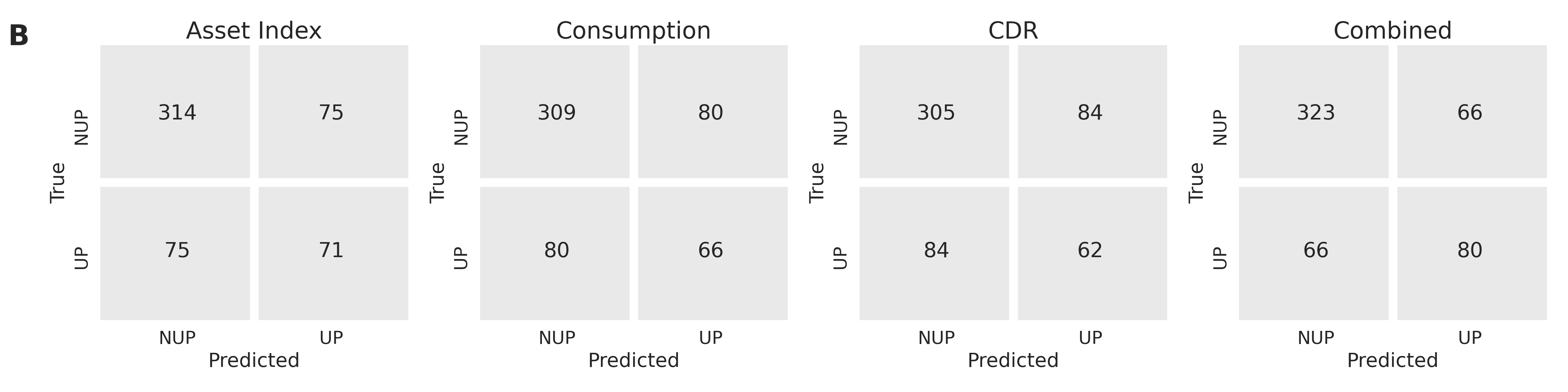}
\includegraphics[width=1\textwidth]{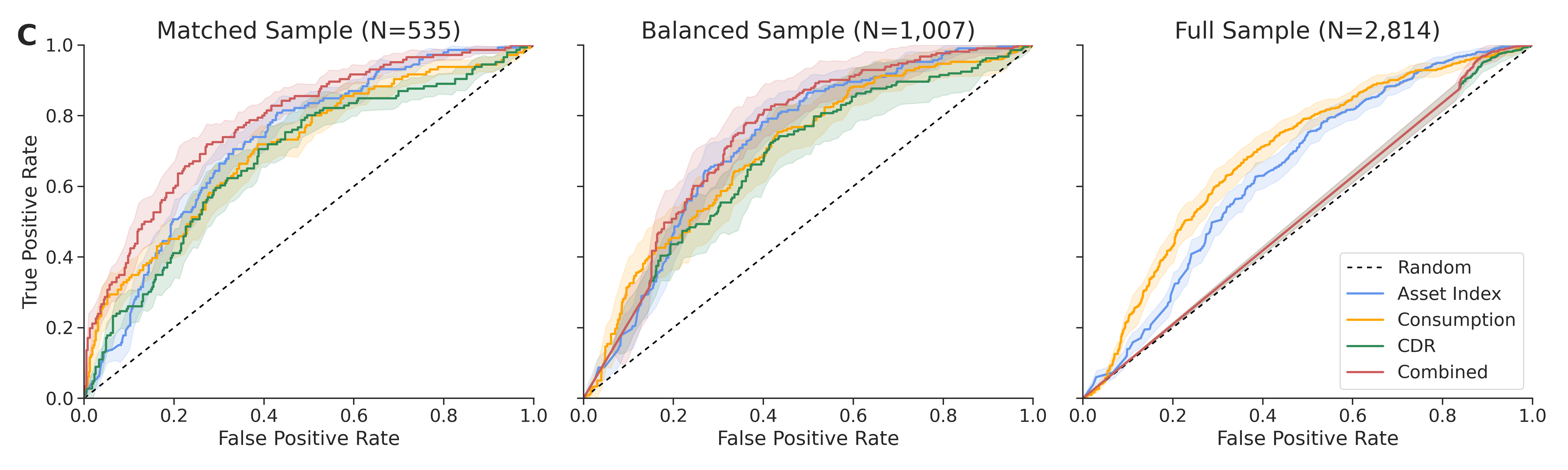}
\caption*{\footnotesize \textit{Notes:} Panel A: Comparing the predictive accuracy of assets, consumption, and CDR-based methods for identifying the ultra-poor in our 535-household sample. To adjust for class balance, thresholds for classification (shown in dashed black vertical lines) are selected such that the correct number of households are identified as ultra-poor. Panel B: Confusion matrices showing the targeting accuracy of each method shown in Panel A. Panel C: ROC curves for each of the four targeting methods. In the third subplot, the CDR-based and combined methods target non-phone-owning households first as described in Section \ref{section:evaluation}}.  
\label{targeting} 
\end{figure}

An overview of these results is provided in Figure \ref{targeting}. Figure \ref{targeting}a shows the distribution of assets and consumption, as well as the distribution of predicted probabilities of being non-ultra-poor generated by the CDR-based and combined methods, separately for the ultra-poor and non-ultra-poor. The dashed vertical line indicates the threshold at which point 27\% of households are classified as ultra-poor; we use this quota because 27\% of households in this sample were designed as ultra-poor by TUP. Figure \ref{targeting}b provides confusion matrices that compare the true status (rows) against the classification made by each method (columns). These confusion matrices are also used to calculate the measures of precision and recall reported in Table \ref{resultstable} Panel A.

We find that the CDR-based method (precision and recall of 42\%) is close in accuracy to methods relying on assets (precision and recall of 49\%) or consumption (precision and recall of 45\%). To evaluate the trade-off between inclusion errors and exclusion errors resulting from selecting alternative cut-off thresholds, Figure \ref{targeting}c shows the ROC curve associated with each classification method, and the associated Area Under the Curve (AUC). AUC scores  are comparable among methods, with assets (AUC=0.73) slightly superior to consumption (AUC=0.71) and the CDR-based method (AUC=0.68).

\subsection{Comparison of Errors Across Methods}
To better understand where the targeting methods are making mistakes, Panel A of Table \ref{tp_vs_fn} indicates how the ultra-poor misclassified as non-ultra-poor (false negatives) compare to the correctly classified ultra-poor (true positives). Panel B shows how the non-ultra-poor misclassified as ultra-poor (false positives) compare to the correctly classified non-ultra-poor (true negatives). 

 Across methods, false negatives score higher on food security, financial inclusion, and psychological well-being than true positives -- that is, all three targeting methods misclassify ultra-poor households as non-ultra-poor when those ultra-poor households are better-off, according to other observable characteristics not used in the targeting.  Likewise, false positives (non-ultra-poor misclassified as ultra-poor) tend to score lower than true negatives across these same measures. The CDR-based method in particular tends to prioritize households that score low on these alternative measures of well-being.

To test for systematic misclassification of certain types of households, Table \ref{mistakesoverlapexclusion} displays the overlap in errors of exclusion and inclusion between methods. Our results suggest that the three classifiers misidentify the same households at a rate only slightly above random.%
\footnote{The rates of overlap should be interpreted relative to the expected overlap in errors for random classifiers. Based on our selection of thresholds such that 27\% of the sample is identified as ultra-poor, our three classifiers misidentify 15-27\% of the non-ultra-poor and 51-65\% of the ultra-poor. If these classifiers were random, we would expect approximately 20\% overlap in inclusion errors and 55\% overlap in exclusion errors.}

\subsection{Combining Targeting Methods}
Since the different targeting methods identify different populations as ultra-poor, there may be complementarities between asset, consumption, and CDR data. As shown in Panel A of Table \ref{resultstable}, we find that a \textit{combined method},  which takes as input the wealth index, total consumption, and the output of the CDR-based method, performs better (AUC = 0.78) than methods using any one data source (AUC = 0.68 - 0.73). As shown in Table~\ref{appendix:combined}, the full method also outperforms methods based on any two data sources (AUC = 0.75 - 0.76). The method that combines CDR and asset data (AUC = 0.76) may, however, be more practical than the combined method, since consumption data is difficult to collect for large populations.

\subsection{Targeting Households Without Phones}
\label{sec:nophones}
An important limitation of CDR-based targeting is that households without phones do not generate CDR. 
Here, we show how targeting performance is impacted when households without phones are included in the analysis. This analysis uses two additional samples of TUP households to evaluate targeting performance: (i) the \textit{balanced sample}, which adds all of the 472 households without phones to the sample of 535 for whom we have matched CDR; the balanced sample is intended to illustrate the performance of CDR-based targeting if CDR were available from all operators in Afghanistan --- though it relies on the assumption that phone-owners observed on our mobile network are representative of all phone owners in Afghanistan (an assumption that is not fully satisfied, as shown in Table \ref{selectionintosample}); and (ii) the \textit{full sample}, which includes all 2,814 households surveyed in the TUP baseline; this sample includes an additional 1,807 households who report owning a phone, but whose number does not match to any number in the CDR provided to us by the single mobile operator.%
\footnote{These 1,807 households include households that report owning a phone on a different network (this network is estimated to have around 30\% market share in Afghanistan), as well as phones on our network that were not active during the six-month period of CDR that we analyze.}

Results in Panels B and C of Table~\ref{resultstable} show the performance of each targeting approach on the balanced and full sample, respectively. Note that as described in Section \ref{section:evaluation}, different targeting quotas are applied for each panel based on the proportion of each sample that is ultra-poor. In the CDR-based and combined approaches, we report performance when the households without CDR are targeted first (i.e. households without CDR are targeted in a random order and then the households predicted to be poorest are targeted until the quota is reached) as well as when households without CDR are targeted last (i.e., after the 535 households with phones are targeted, households without phones are included in a random order until the quota is reached).

Unsurprisingly, these results suggest that CDR-based targeting is not effective when a large portion of the target population does not own a phone (e.g. Panel C of Table~\ref{resultstable}, where only 16\% of the sample has matching CDR). However, when we simulate more realistic levels of phone ownership in Panel B (84\% of the households, based on our survey data), CDR-based targeting is once again comparable to asset- or expenditure-based targeting, particularly when households without phones are targeted first (AUC = 0.72, 0.70, 0.68 for assets, consumption, and CDR, respectively). On the other hand, if households without phones are targeted last (for example, if program administrators base targeting wholly on CDR and provide no benefits to any household without a phone), the CDR-based method only improves marginally on random targeting.

\subsection{Additional tests and simulations} 

Our main analysis considers the household head to be the unit of analysis. As described in Section~\ref{section:cdr}, this analysis is based on matching survey-based indices to phone data from the household head, which is consistent with the design of the TUP program and the TUP survey sample frame. An alternative approach matches survey data reported by the household head to all phone numbers associated with the household. As shown in Table~\ref{appendix:individual}, the predictive accuracy of these models is slightly attenuated relative to the benchmark results (Table~\ref{appendix:mldetails}).

We also explore the extent to which CDR can be used to predict other measures of socioeconomic status. Our main analysis focuses on the household's ultra-poor designation as the ground truth measure of poverty, since this label was both carefully curated and the actual criterion used to determine TUP eligibility. In Table~\ref{appendix:mlresults}, we report the accuracy with which CDR (obtained from the household head, who is typically male) can predict consumption and asset-based wealth (elicited from the primary woman of each household).%
\footnote{Due to the design of the TUP survey, which interviewed women in the household, we cannot avoid this mismatch between the survey respondent and the phone owner.}
In general, these machine learning models trained to directly predict consumption or asset-based wealth do not perform well. This result contrasts with prior work documenting the predictive ability of CDR for measuring asset-based wealth \citep[e.g.][]{blumenstock2015}. We suspect a key difference in our setting -- aside from the fact that we are matching CDR to socioeconomic status at the \textit{household} rather than the \textit{individual} level -- is the homogeneity of the beneficiary population: whereas \citet{blumenstock2015} uses machine learning to predict the wealth of a nationally-representative sample of Rwandan phone owners, our sample consists of 535 individuals from the poorest villages of a single province in Afghanistan, where even the relatively wealthy households are quite poor.

\section{Discussion}
\label{sec:discussion}
Our key finding is that, in a sample of 535 phone-owning households in poor villages in Afghanistan, machine learning methods leveraging phone data are nearly as accurate at identifying ultra-poor households as standard asset- and consumption-based methods. Further, we find that methods combining survey data with CDR perform better than methods using a single data source.  However, as we demonstrate empirically, low rates of phone ownership --- or the inability to access data from all operators --- can undermine the value of CDR-based targeting. In our setting, the CDR-based approach still works well if households without phones are targeted before the CDR-based algorithm selects the poorest households with phones. However, this approach may not be appropriate in other contexts where phone ownership is less predictive of wealth, or where potential beneficiaries have the ability to strategically under-report phone ownership \citep{bjorkegren2020manipulation}.

As mobile phone penetration rates continue to rise in LMICs \citep{gsma}, and as programs increasingly rely on mobile phones and money to distribute benefits \citep[cf.][]{gentilini_social_2020}, CDR-based targeting methods will likely play a more prominent role in the set of options considered by policymakers and program administrators --- particularly in contexts like Afghanistan, where traditional targeting benchmarks are missing or of low quality. In just the past few years, for instance, data from mobile phone operators was used in the design of social assistance programs in Colombia, the Democratic Republic of the Congo, Pakistan, and Togo \citep{gentilini_social_2020,gentilini2021cash,aiken_machine_2021}. We conclude by highlighting a few policy important considerations for CDR-based targeting.

\paragraph{Speed and cost} An advantage of CDR-based targeting is that it can be used in contexts where face-to-face contact is not feasible, dramatically reducing the time required to implement a targeted  program. While it typically takes many months (or years) to implement a proxy-means test (PMT), community-based targeting (CBT), or consumption-based targeting, a CDR-based model can be trained in just a few weeks (see Appendix \ref{sec:appendix:costs}). Likewise, the marginal costs per household screened are substantially lower with CDR-based targeting than with CBT, PMT, or consumption-based targeting. For instance, Table \ref{appendix:costs} uses cost estimates obtained from the literature (and detailed in Table \ref{appendix:costliterature}) to estimate targeting costs for the TUP program.\footnote{In our cost calculations we obtain estimates for a CBT, rather than the hybrid approach used in the TUP program, as there is more information available on CBT-only costs in the literature. However, as the CBT cost can be interpreted as a lower bound for the cost of a hybrid approach, our qualitative results also apply to a hybrid approach.}  Whereas the marginal costs of screening an individual with a CBT or PMT are estimated at \$2.20 and \$4.00, respectively, the marginal cost of screening with CDR is negligible (see Appendix \ref{sec:appendix:costs}).\footnote{Marginal costs of CDR-based targeting are negligible because we assume no contact with screened individuals is required. In practice, it may be desirable to solicit informed consent to access CDR. If consent were collected in-person, the marginal costs would approach that of a PMT; if collected over the phone,  there would still be significant cost savings, see Appendix Section \ref{sec:appendix:costs}).}
For the entire TUP program, which screened around 125,721 households in six provinces, CBT and PMT would add an additional estimated \$276,586 and \$502,884, respectively, corresponding to 2.18\% and 3.97\% of the total program budget. 

\paragraph{Data access and privacy} Access to phone data is necessary for CDR-based targeting. As we show, targeting performance degrades considerably when CDR are not available for subsets of the population. Encouragingly, the past several years have been characterized by a trend towards public sector access to CDR, particularly in the context of the COVID-19 pandemic, during which mobile network operators shared CDR with governments, researchers, and NGOs for social protection purposes \citep[cf.][]{gentilini_social_2020,gentilini2021cash,aiken_machine_2021}. CDR have also been shared with the public sector for public health and humanitarian aid applications \citep{milusheva2021challenges}. Access issues aside, CDR contain private and sensitive data, including phone numbers and location traces. While much has been written about enabling responsible use of CDR for humanitarian response \citep[e.g.][]{de2018privacy,oliver2020mobile}, to date no consistent privacy standards exist. Informed consent can increase participant agency, but also complicates the implementation logistics. Likewise, there may be ways to incorporate differential privacy or other privacy enhancing technologies into a CDR-based targeting system, but such privatization would likely decrease targeting accuracy \citep{hu2015differential}.

\paragraph{Algorithmic transparency and strategic behavior} Using CDR to determine program eligibility may introduce incentives for people to manipulate their phone use. This consideration is not unique to CDR, as degrees of manipulation have been documented in social programs that use proxy means tests and other traditional targeting mechanisms \citep{camacho_manipulation_2011,banerjee_lack_2018}. Indeed, complex and non-linear machine learning algorithms, like the one presented in this paper, may obfuscate the logic behind targeting decisions, reducing the scope for manipulation. However, society often demands transparency in algorithmic decision-making, as black-box decisions are difficult to audit or hold to account. There is therefore a tension between the goals of increasing transparency and reducing manipulation, though recent advances in machine learning explore mechanisms for pursuing both objectives at once \citep{bjorkegren2020manipulation}.


\paragraph{Centralized vs. local knowledge} CDR-based methods enable a top-down, centralized and standardized approach to program targeting, rather than a bottom-up approach that prioritizes local knowledge that can be elicited, for example, through community wealth rankings. While the empirical results in this paper indicate that the efficiency gains from CDR-based targeting are substantial, it may reinforce existing power structures \citep{taylor,blumenstockbigdata,abebe_narratives_2021}. Efficiency gains should also be considered within the context of evidence suggesting that participating communities may prefer community-based approaches \citep{alatas2012}, but also may perceive them as less legitimate \citep{Premandetal2020}.

\paragraph{} To summarize, our results suggest that there is potential for using CDR-based methods to determine eligibility for economic aid or interventions, substantially reducing program targeting overhead and costs. Our results also indicate that CDR-based methods may complement and enhance existing survey-based methods. We note, however, that the practical and ethical limitations to CDR-based targeting are significant. We emphasize the need to consider these limitations and the constraints of specific local contexts alongside the efficiency gains offered by CDR-based targeting.

\pagebreak

\bibliographystyle{apacite}
\bibliography{main}

\clearpage
\pagebreak
\section*{Tables and Figures}

\begin{table}[htb] \centering 
  \caption{Summary statistics for different samples of survey respondents\label{selectionintosample}}
\renewcommand{\arraystretch}{1.2}
\begin{threeparttable}
\begin{footnotesize}
\begin{tabular}{lcccc}
\toprule
 & (1) & (2) & (3) & (4) \\
  \textbf{Outcome}  & \textbf{Full sample} & \textbf{Matched} & \textbf{Unmatched} & \textbf{Unmatched} \\
                                    & \textbf{(all observations)} & \textbf{Subsample} & \textbf{Owns Phone} & \textbf{No Phone} \\ 
\midrule 
\\ [-0.8em]
\multicolumn{5}{l}{\textit{Panel A: Balance of Covariates}} \\ 
[0.3em]
                        Ultra-Poor &      0.42 (0.49) &       0.27 (0.45) &            0.40 (0.49) &         0.66 (0.47) \\
                       Asset Index &      0.00 (2.01) &       1.36 (2.60) &           -0.05 (1.76) &        -1.35 (0.79) \\
                  Log Expenditures &      4.43 (0.71) &       4.64 (0.70) &            4.46 (0.70) &         4.12 (0.65) \\
                          \# Phones &      1.35 (1.18) &       1.72 (1.33) &            1.59 (1.04) &         0.00 (0.00) \\
               Food Security Index &      0.30 (0.90) &       0.35 (0.74) &            0.34 (0.93) &         0.10 (0.89) \\
         Financial Inclusion Index &      0.15 (1.27) &       0.34 (1.39) &            0.15 (1.32) &        -0.05 (0.79) \\
    Psychological Well-being Index &      0.35 (1.01) &       0.38 (1.00) &            0.43 (0.97) &        -0.02 (1.07) \\
                         CWR Group &      0.62 (0.90) &       0.89 (1.02) &            0.62 (0.88) &         0.26 (0.66) \\
\\ [-0.5em]
\multicolumn{5}{l}{\textit{Panel B: Correlations Between Outcomes}} \\
[0.3em]
       Ultra-Poor $\longleftrightarrow$ Asset Index &            -0.32 &             -0.30 &                  -0.27 &               -0.14 \\
  Ultra-Poor $\longleftrightarrow$ Consumption &            -0.39 &             -0.30 &                  -0.39 &               -0.26 \\
 Asset Index $\longleftrightarrow$ Consumption &             0.37 &              0.34 &                   0.34 &                0.15 \\ 
 \\[-0.9em]
                                 $N$ &             2,814 &               535 &                   1,807 &                 472 \\
\bottomrule
 \\[-2.5ex] 
\end{tabular} 
\end{footnotesize}
\begin{tablenotes}[normal,flushleft]
\footnotesize
\item \emph{Notes}: Table reports average characteristics, with standard deviations in parentheses, of TUP survey respondents. Each column represents a different sample of respondents: (1) all respondents in the TUP survey; (2) Just those respondents who own a phone, where the phone number matches to the CDR obtained from the mobile phone operator; (3) Respondents who report owning a phone, but whose phone number does not match to the CDR obtained from the operator; (4) Respondents who report they do not own a phone.
\end{tablenotes}
\end{threeparttable}
\end{table} 

\pagebreak
\begin{table}[H] \centering 
  \caption{Targeting simulation results\label{resultstable}}
\renewcommand{\arraystretch}{1.2}
\begin{threeparttable}
\begin{footnotesize}
\begin{tabular}{lcccc}
\toprule
 & (1) & (2) & (3) & (4) \\
 \textbf{Targeting Method} &
 \multicolumn{1}{c}{\textbf{AUC}} & 
 \multicolumn{1}{c}{\textbf{Accuracy}} & 
 \multicolumn{1}{c}{\textbf{Precision}} & 
 \multicolumn{1}{c}{\textbf{Recall}} \\
\midrule
\\[-.8em]
\multicolumn{5}{l}{\emph{Panel A: Matched Sample (N=535) - for whom we have survey and CDR data}} \\[0.3em]
                              Random &  0.50 (0.028) &  0.60 (0.025) &  0.27 (0.038) &  0.27 (0.038) \\
                           Asset Index &  0.73 (0.024) &  0.72 (0.020) &  0.49 (0.041) &  0.49 (0.041) \\
                      Consumption &  0.71 (0.026) &  0.69 (0.023) &  0.45 (0.038) &  0.45 (0.038) \\
                                   CDR &  0.68 (0.027) &  0.69 (0.021) &  0.42 (0.042) &  0.42 (0.042) \\
                              Combined &  0.78 (0.022) &  0.75 (0.020) &  0.55 (0.039) &  0.55 (0.039) \\

\\ [-0.8em]
\multicolumn{5}{l}{\emph{Panel B: Balanced Sample (N=1,007) - as above, plus households without phones}} \\[0.3em]
                              Random &  0.50 (0.017) &  0.90 (0.006) &  0.05 (0.010) &  0.05 (0.010) \\
                           Asset Index &  0.72 (0.026) &  0.90 (0.006) &  0.10 (0.013) &  0.10 (0.013) \\
                      Consumption &  0.70 (0.028) &  0.90 (0.006) &  0.15 (0.025) &  0.15 (0.025) \\
                                CDR (Target Phoneless First) &  0.68 (0.030) &  0.90 (0.006) &  0.11 (0.035) &  0.11 (0.035) \\
           CDR (Target Phoneless Last) &  0.51 (0.028) &  0.90 (0.006) &  0.12 (0.033) &  0.12 (0.033) \\

               Combined (Target Phoneless First) &  0.74 (0.026) &  0.90 (0.006) &  0.11 (0.046) &  0.11 (0.046) \\
      Combined (Target Phoneless Last) &  0.57 (0.022) &  0.90 (0.006) &  0.18 (0.007) &  0.18 (0.007) \\
\\[-0.8em]
\multicolumn{5}{l}{\emph{Panel C: Full Sample (N=2,814) - as above, plus households with phones on other networks}} \\[0.3em]
                              Random &  0.50 (0.009) &  0.89 (0.005) &  0.06 (0.007) &  0.06 (0.007) \\
                           Asset Index &  0.65 (0.017) &  0.89 (0.005) &  0.07 (0.014) &  0.07 (0.014) \\
                      Consumption &  0.69 (0.015) &  0.89 (0.006) &  0.08 (0.031) &  0.08 (0.031) \\
                                CDR (Target Phoneless First) &  0.52 (0.008) &  0.89 (0.005) &  0.06 (0.008) &  0.06 (0.008) \\
           CDR (Target Phoneless Last) &  0.48 (0.008) &  0.89 (0.005) &  0.08 (0.010) &  0.08 (0.010) \\

               Combined (Target Phoneless First) &  0.52 (0.008) &  0.89 (0.005) &  0.06 (0.008) &  0.06 (0.008) \\
      Combined (Target Phoneless Last) &  0.49 (0.008) &  0.89 (0.005) &  0.09 (0.009) &  0.09 (0.009) \\
\bottomrule
 \\[-2.5ex] 
\end{tabular} 
\end{footnotesize}
\begin{tablenotes}[normal,flushleft]
\footnotesize
\item \emph{Notes}: Four different measures of performance (columns) reported for different targeting methods (rows), using different samples of survey respondents (panels). Standard deviations, calculated using 1,000 bootstrap samples, in parentheses. Panel A: The 535-household subsample that is matched to CDR. Panel B: The 535-household matched sample, plus the 472 households that do not have a phone; this is meant to approximate targeting performance if CDR from all mobile networks were available. Sample weights are applied as described in Section \ref{section:evaluation}. Panel C: All 2,814 observations from the TUP survey, including households matched to CDR, households that own phones not matched to CDR, and households without phones, with sample weights applied. For Panels B and C, we simulate two types of CDR-based targeting: targeting households without phones first and targeting households without phones last. 
\end{tablenotes}
\end{threeparttable}
\end{table}

\begin{table}[H] \centering 
  \caption{What types of households are misclassified? \label{tp_vs_fn}}
\renewcommand{\arraystretch}{1.2}
\begin{threeparttable}
\begin{footnotesize}
\begin{tabular}{p{3cm}p{1cm}p{1cm}p{1cm}p{1cm}p{1cm}p{1cm}p{1cm}p{1cm}p{1cm}p{1cm}}
\toprule
\multicolumn{10}{l}{\emph{Panel A: Ultra-Poor Households (Differences Between True Positives and False Negatives)}} \\ [.5em]

& \multicolumn{3}{c}{\textbf{Asset Index}} & \multicolumn{3}{c}{\textbf{Consumption}} & \multicolumn{3}{c}{\textbf{CDR}} \\

& \textbf{TP} &  \textbf{FN} &  \textbf{Diff.} & \textbf{TP} &  \textbf{FN} &  \textbf{Diff.} & \textbf{TP} &  \textbf{FN} &  \textbf{Diff.} \\
\cmidrule{2-10}
      Ultra-Poor &   1.00 (0.00) &   1.00 (0.00) &   0.00 (0.00) &   1.00 (0.00) &   1.00 (0.00) &   0.00 (0.00) &   1.00 (0.00) &   1.00 (0.00) &   0.00 (0.00) \\
                   Asset Index &  -1.03 (0.49) &   1.18 (1.34) &  -2.21 (0.17) &  -0.34 (1.09) &   0.47 (1.69) &  -0.81 (0.23) &  -0.09 (1.16) &   0.25 (1.70) &  -0.34 (0.24) \\
              Consumption &   4.21 (0.70) &   4.40 (0.62) &  -0.19 (0.11) &   3.78 (0.32) &   4.74 (0.56) &  -0.96 (0.07) &   4.29 (0.60) &   4.32 (0.71) &  -0.02 (0.11) \\
                      \# Phones &   0.89 (0.68) &   1.63 (1.12) &  -0.74 (0.15) &   1.02 (0.73) &   1.48 (1.14) &  -0.46 (0.16) &   1.18 (0.61) &   1.33 (1.21) &  -0.16 (0.15) \\
           Food Security Index &  -0.59 (1.13) &  -0.51 (1.10) &  -0.08 (0.18) &  -0.83 (1.19) &  -0.32 (0.99) &  -0.51 (0.18) &  -0.51 (1.14) &  -0.58 (1.09) &   0.07 (0.19) \\
     Financial Inclusion Index &  -0.00 (0.79) &   0.29 (1.02) &  -0.29 (0.15) &   0.10 (0.80) &   0.19 (1.02) &  -0.09 (0.15) &   0.16 (0.98) &   0.14 (0.88) &   0.02 (0.16) \\
 Psychological Wellbeing Index &  -0.35 (0.92) &  -0.13 (0.94) &  -0.22 (0.15) &  -0.37 (0.86) &  -0.12 (0.98) &  -0.24 (0.15) &  -0.31 (0.81) &  -0.17 (1.02) &  -0.14 (0.15) \\
                     CWR Group &   0.09 (0.44) &   0.01 (0.12) &   0.07 (0.05) &   0.02 (0.12) &   0.08 (0.41) &  -0.06 (0.05) &   0.06 (0.40) &   0.04 (0.24) &   0.03 (0.06) \\[1.2em]
\midrule
\\[-.5em]
\multicolumn{10}{l}{\emph{Panel B: Non-Ultra-Poor Households (Differences Between True Negatives and False Positives)}}\\[.5em]

& \multicolumn{3}{c}{\textbf{Asset Index}} & \multicolumn{3}{c}{\textbf{Consumption}} & \multicolumn{3}{c}{\textbf{CDR}} \\

& \textbf{TN} &  \textbf{FP} &  \textbf{Diff.} & \textbf{TN} &  \textbf{FP} &  \textbf{Diff.} & \textbf{TN} &  \textbf{FP} &  \textbf{Diff.} \\
\cmidrule{2-10}
Ultra-Poor &   0.00 (0.00) &   0.00 (0.00) &   0.00 (0.00) &   0.00 (0.00) &   0.00 (0.00) &   0.00 (0.00) &   0.00 (0.00) &   0.00 (0.00) &   0.00 (0.00) \\
                   Asset Index &   2.53 (2.62) &  -1.08 (0.50) &   3.61 (0.16) &   2.06 (2.92) &   0.94 (1.75) &   1.12 (0.26) &   1.94 (2.87) &   1.43 (2.27) &   0.51 (0.30) \\
              Consumption &   4.82 (0.66) &   4.57 (0.65) &   0.25 (0.08) &   4.97 (0.58) &   3.98 (0.23) &   0.99 (0.04) &   4.78 (0.68) &   4.74 (0.61) &   0.04 (0.08) \\
                      \# Phones &   2.11 (1.43) &   0.96 (0.76) &   1.15 (0.12) &   1.98 (1.49) &   1.52 (0.92) &   0.46 (0.13) &   1.91 (1.44) &   1.80 (1.24) &   0.11 (0.16) \\
           Food Security Index &   0.24 (0.87) &  -0.16 (1.03) &   0.40 (0.13) &   0.24 (0.88) &  -0.14 (0.99) &   0.37 (0.12) &   0.15 (0.91) &   0.18 (0.94) &  -0.02 (0.12) \\
     Financial Inclusion Index &   0.80 (4.92) &  -0.01 (0.82) &   0.82 (0.29) &   0.77 (4.94) &   0.18 (1.24) &   0.59 (0.31) &   0.78 (4.98) &   0.17 (1.10) &   0.61 (0.31) \\
 Psychological Wellbeing Index &   0.69 (0.97) &   0.21 (0.75) &   0.47 (0.10) &   0.62 (0.98) &   0.49 (0.80) &   0.13 (0.11) &   0.62 (0.95) &   0.49 (0.93) &   0.13 (0.12) \\
                     CWR Group &   1.30 (1.00) &   0.84 (0.96) &   0.46 (0.12) &   1.23 (1.03) &   1.13 (0.94) &   0.10 (0.12) &   1.26 (1.01) &   1.01 (0.98) &   0.25 (0.12) \\
\bottomrule
 \\[-2.5ex] 
\end{tabular} 
\end{footnotesize}
\begin{tablenotes}[normal,flushleft]
\footnotesize
\item \emph{Notes}: Table shows the average characteristics (with standard deviations in parentheses) of households that are correctly classified (True Positives [TP] and True Negatives [TN]) and incorrectly classified (False Negatives [FN] and False Positives [FP]), as well as the difference in average characteristics between correctly and incorrectly classified households (Diff.).  Panel A: Differences between ultra-poor households correctly classified as such and those misclassified as non-ultra-poor (errors of exclusion). Panel B: Differences between non-ultra-poor households correctly classified as such and those misclassified as ultra-poor (errors of inclusion).
\end{tablenotes}
\end{threeparttable}
\end{table} 

\begin{table}[H] \centering 
  \caption{Overlap in targeting errors between methods\label{mistakesoverlapexclusion}}
\renewcommand{\arraystretch}{1.2}
\begin{threeparttable}
\begin{footnotesize}
\begin{tabular}{lcccc}
\toprule
                  & \textbf{Asset Index} & \textbf{Consumption} &      \textbf{CDR} & \textbf{Combined} \\
\midrule
\\ [-.9em]
\multicolumn{5}{l}{\emph{Panel A: Overlap in Errors of Exclusion}} \\
      Asset Index &     100.00\% &           65.33\% &   57.33\% &   66.67\% \\
Consumption &      61.25\% &          100.00\% &   56.25\% &   62.50\% \\
              CDR &      51.19\% &           53.57\% &  100.00\% &   63.10\% \\
         Combined &      75.76\% &           75.76\% &   80.30\% &  100.00\% \\
\\[-.8em]
\multicolumn{5}{l}{\emph{Panel B: Overlap in Errors of Inclusion}} \\
      Asset Index &     100.00\% &           26.67\% &   22.67\% &   48.00\% \\
Consumption &      25.00\% &          100.00\% &   16.25\% &   37.50\% \\
              CDR &      20.24\% &           15.48\% &  100.00\% &   46.43\% \\
         Combined &      54.55\% &           45.45\% &   59.09\% &  100.00\% \\
\bottomrule
 \\[-2.5ex] 
\end{tabular} 
\end{footnotesize}
\begin{tablenotes}[normal,flushleft]
\footnotesize
\item \emph{Notes}: Table measures the extent to which the targeting errors produced by each pair of targeting methods overlap. Evaluation is performed on the matched sample of 535 TUP respondents. Panel A: Overlap between ultra-poor households that are misclassified as non-ultra-poor (errors of exclusion) for each targeting method. Panel B: Overlap between non-ultra-poor households that are misclassified as ultra-poor (errors of inclusion).
\end{tablenotes}
\end{threeparttable}
\end{table} 

\newpage
\begin{center}
  \huge{Online Appendix}
\end{center}
\beginsupplement
\appendix


\section{Machine learning methods and hyperparameters}
\label{appendix:ml}

Although our paper is focused on identifying the ultra-poor with CDR, we experiment with predicting four measures of ground-truth welfare with CDR features: ultra-poor status (binary), below the national poverty line (binary), asset index (continuous), and log consumption (continuous). For the binary measures, we experiment with four classification models: logistic regression (unregularized), logistic regression with L1 penalty, a random forest, and a gradient boosting model. For the continuous measures, we experiment with four regression models: linear regression, LASSO regression, a random forest, and a gradient boosting model. The linear models and random forest are implemented in Python’s scikit-learn package. The gradient boosting model is implemented with Microsoft’s LightGBM. 

In each case, we produce predictions out-of-sample over 10-fold cross validation. We use nested cross-validation to tune the hyperparameters of each model over 5-fold cross-validation within each of the outer folds to avoid any information leakage between folds. We report both the mean score across the 10 folds as well as the overall score when data from all folds is pooled together. For the linear models and random forest, missing data is mean-imputed and each feature is scaled to zero mean and unit variance before fitting models (these transformations are done separately for each fold, with parameters fitted only on the training data for each fold). For the gradient boosting model missing values are left as-is and features are not scaled. We re-fit the model on the entire data, again tuning hyperparameters over 5-fold cross validation, to report selected hyperparameters and feature importances. We also report the top 5 features for each model, determined by the magnitude of the coefficient for the linear models, and by and by maximum impurity reductions for the tree-based models. 

Hyperparameters are selected from the following grids for each model:

\paragraph{Linear/Logistic Regression}
\begin{itemize}
    \item Drop columns with missingness over: \{50\%, 80\%, 100\%\}
    \item Drop columns with variance under: \{0, 0.01, 0.1\}
    \item Winsorization limit: \{0\%, 1\%, 5\%\}
\end{itemize}

\paragraph{LASSO Regression}
\begin{itemize}
    \item Drop columns with missingness over: \{50\%, 80\%, 100\%\}
    \item Drop columns with variance under: \{0, 0.01, 0.1\}
    \item Winsorization limit: \{0\%, 1\%, 5\%\}
    \item L1 penalty: \{0.00001, 0.0001, 0.001, 0.01, 0.1, 1, 10, 100\}
\end{itemize}

\paragraph{Random Forest}
\begin{itemize}
    \item Drop columns with missingness over: \{50\%, 80\%, 100\%\}
    \item Drop columns with variance under: \{0, 0.01, 0.1\}
    \item Winsorization limit: \{0\%, 1\%, 5\%\}
    \item Number of Trees: \{20, 50, 100\}
    \item Maximum Depth: \{1, 2, 4, 6, 8, 10, 12\}
\end{itemize}

\paragraph{Gradient Boosting Model}
\begin{itemize}
    \item Drop columns with missingness over: \{50\%, 80\%, 100\%\}
    \item Drop columns with variance under: \{0, 0.01, 0.1\}
    \item Winsorization limit: \{0\%, 1\%, 5\%\}
    \item Number of Trees: \{20, 50, 100\}
    \item Minimum data in leaf: \{5, 10\}
    \item Number of leaves: \{5, 10, 20\}
    \item Learning rate: \{0.05, 0.075\}
\end{itemize}

\pagebreak
\section{Abbreviations in Feature Names}
\label{appendix:abbreviations}

\begin{figure}[H]
    \centering
    \includegraphics[width=1\textwidth]{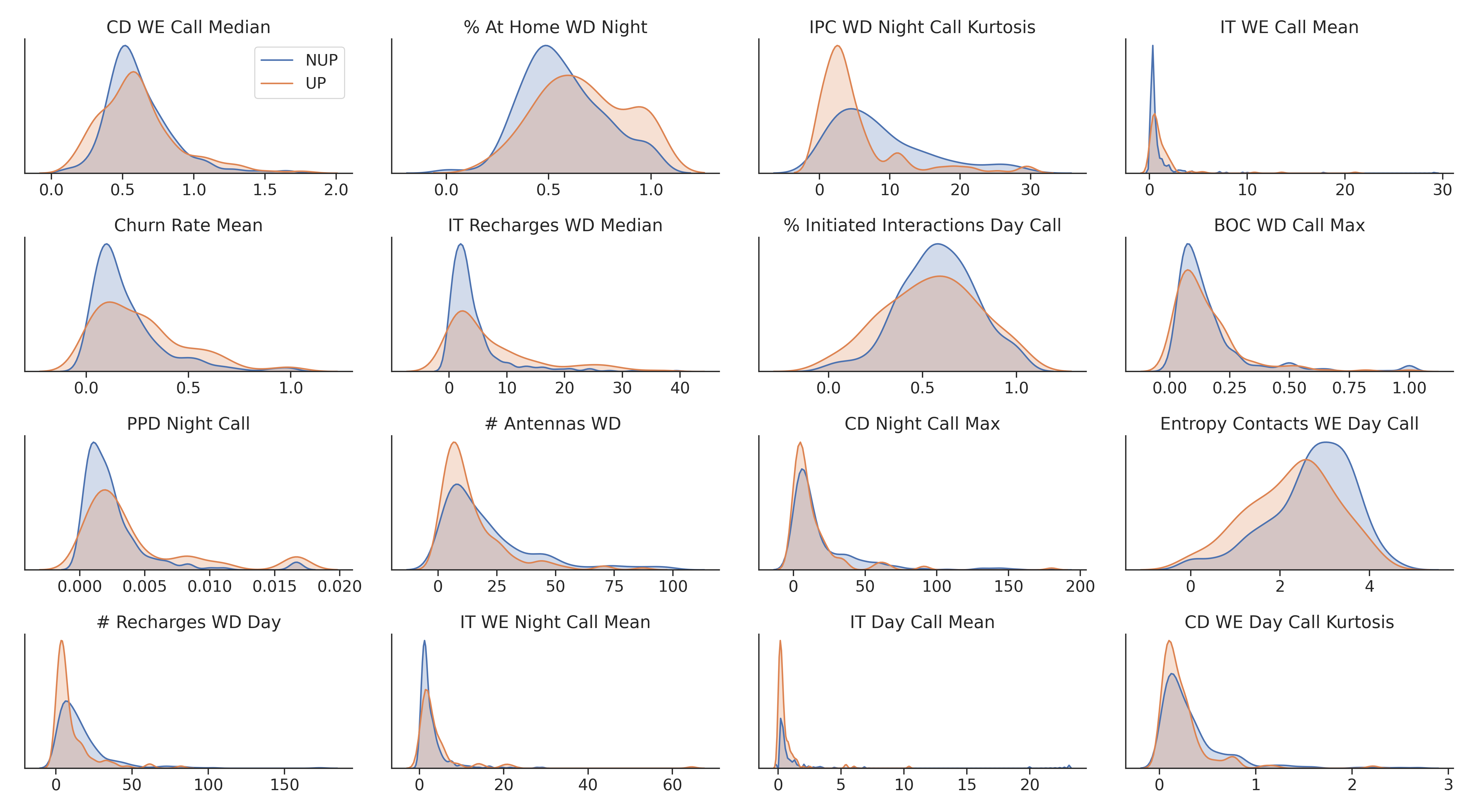}
    \caption{Kernel density estimates for 16 of the most important features for predicting ultra-poor status from CDR, with density estimates shown separately from UP and NUP households. Since many features are near-redundant, rather than showing the raw top 16 features from the table above, we show 16 selected features from the top 50.  See Appendix \ref{appendix:abbreviations} for abbreviations in feature names.}
    \label{appendix:feature_kdes}
\end{figure}

Figure \ref{appendix:feature_kdes} and Tables \ref{appendix:mlresults}, \ref{appendix:importances}, and \ref{appendix:individual} use a set of abbreviations in CDR feature names. This appendix lists the relevant abbreviations.

\begin{itemize}
    \item BOC: Balance of contacts
    \item CD: Call duration
    \item IPC: Interactions per contact
    \item IT: Interevent time
    \item NOI: Number of interactions
    \item PPD: Percent pareto durations (percentage of call contacts accounting for 80\% of call time)
    \item PPI: Percent pareto interactions (percentage of contacts accounting for 80\% of subscriber's interactions)
    \item RD: Response delay
    \item RR: Response rate
    \item WD: Weekday 
    \item WE: Weekend
\end{itemize}

\section{Cost and Speed Calculations}
\label{sec:appendix:costs}
In the discussion section we provide a cost and speed comparison between targeting methods, as some of the value-add of the phone-based targeting approach relies on how cheap and quick it is compared to asset, consumption- or CBT-based targeting approaches. Administrative data on targeting costs was not collected as part of the TUP program, so we turn to other studies of program targeting to estimate the costs of CBT and asset-based (or PMT) methods. We treat the costs of an asset index-based and PMT approach as equivalent in this section, as they both require comprehensive household surveys.\footnote{In practice, an asset-based approach may be slightly cheaper than a PMT, as it does not require conducting a consumption module for a subset of surveys to train a PMT.} We identify three studies that provide variable targeting costs for PMT and CBT methods: \cite{alatas2012} provide variable costs for CBT and PMT-based targeting of a single program in Indonesia; \cite{karlan2013} provide variable costs for CBT and PMT-based targeting of an ultra-poor program in Honduras and one in Peru; and \cite{schnitzer2021targeting} provide variable costs for three CBT-based programs and four PMT-based programs in seven countries in Sub-Saharan Africa.\footnote{To our knowledge, no studies incorporate fixed targeting costs, as these are typically indistinguishable from fixed costs of other components of program set-up.} Table \ref{appendix:costliterature} summarizes the cost estimates from each of these papers; we use the median per-household targeting cost for each method in our analysis (\$2.20 per household for CBT and \$4.00 per household for PMT). While using these global estimates to inform our model of targeting costs in Afghanistan is not ideal, since no data on targeting costs from the TUP program or other anti-poverty programs is available for the country, these values are the best available estimate on which to base our cost analysis. 

We are unable to find any papers that document the targeting cost associated with consumption-based targeting, as consumption data is rarely used as a real-world targeting strategy. We therefore consider the costs of consumption to be strictly greater than the costs of targeting on a PMT, since consumption modules take longer to collect than PMT data in household surveys. In practice, we expect that the cost of targeting on consumption would be substantially greater than the cost of targeting on a PMT.

For phone-based targeting, we associate no cost with the collection and analysis of phone data. While in some cases phone data may require purchase from the operator, partnerships between mobile network operators and governments for social protection and public health applications have not, to date, involved payment \citep{milusheva2021challenges}. The fixed cost of mobile data analysis is non-negligible but its contribution to marginal cost is close to zero as the number of screened households increases. A phone-based targeting method that collects informed consent from program applicants to analyze phone data would have nonzero marginal costs, though the cost of consent would depend on the modality of consent collection. If consent was collected in person, these costs would likely be only slightly lower than those of a PMT, as every household would need to be surveyed in person. If consent was collected over the phone via SMS or voice, these costs would likely  be significantly lower.  

It is worth noting that our benchmark in this paper is the hybrid model with a CBT plus verification component, but due to limited estimates in the literature we leave this strategy out of our cost analysis. We consider the CBT a lower-bound estimate for the hybrid strategy, and therefore our results would be qualitatively unchanged if the hybrid strategy were also considered in cost comparison. \cite{alatas2012} suggest that there are synergies in targeting approaches so that combining approaches is less costly than the sum of the costs of the two approaches individually, but costs are certainly greater than that of CBT targeting alone.  

Our cost analysis finally relies upon administrative data from the TUP program. The TUP program in its entirety served 7,500 households across six provinces of Afghanistan. While there is no data available on the total number of households screened by the TUP program, the portion of the program in Balkh province that was enrolled in the RCT identified 1,235 ultra-poor households out of 20,702 households screened \citep{bedoya2019}. Assuming similar eligibility rates across Afghanistan, we estimate that the TUP program as a whole likely screened around 125,721 households. We use this value to estimate total targeting costs for the TUP program under counterfactual targeting approaches. Eligible households received benefits totaling \$1,688, including a productive asset, cash transfers, a health voucher, training, biweekly social worker visits, and veterinarian visit once every two months during the year of intervention. The total benefits dispersed by the program were therefore on the order of 12.7 million (although the total program costs, including overhead, were closer to 15 million \citep{bedoya2019}); we use the total value of benefits to compare the costs of program targeting using our set of counterfactual targeting approaches to the direct costs of program benefits in Table \ref{appendix:costs}. We find that targeting costs for a PMT or asset-based approach would represent approximately 3.97\% of the total benefits delivered in the program; costs for a CBT approach would represent approximately 2.18\% of the total benefits. In comparison, costs for the phone-based approach would be negligible.

When it comes to speed, in-person data collection for an asset-based (or PMT) targeting approach typically takes months or years to prepare and implement \citep{WorldBank_2020}. The CDR-based approach can be rolled out comparatively quickly --- but there are still practical hurdles to implementation. First, training data for the CDR-based poverty prediction model must be collected, preferably shortly prior to program roll-out \citep{aiken_machine_2021}. While in the TUP project training data was collected in-person in a household survey, in other contexts training data collection was expedited via a phone survey \citep{blumenstock2015, aiken_machine_2021}. Even if data is collected over the phone, it will typically take several weeks to design a survey instrument and collect data. Second, the CDR-based method requires data from mobile network operators. Data sharing agreements with mobile network operators take at minimum a few weeks to arrange, and substantially longer in the worst case \citep{milusheva2021challenges}. Third, and finally, training a CDR-based poverty prediction model is expensive in terms of memory,  computing power, and human capacity, and will likely take several weeks to implement.

\pagebreak
\section*{Supplementary Tables and Figures}
\label{appendix:supplementaryfigures}


\begin{figure}[H]
    \centering
    \includegraphics[width=1\textwidth]{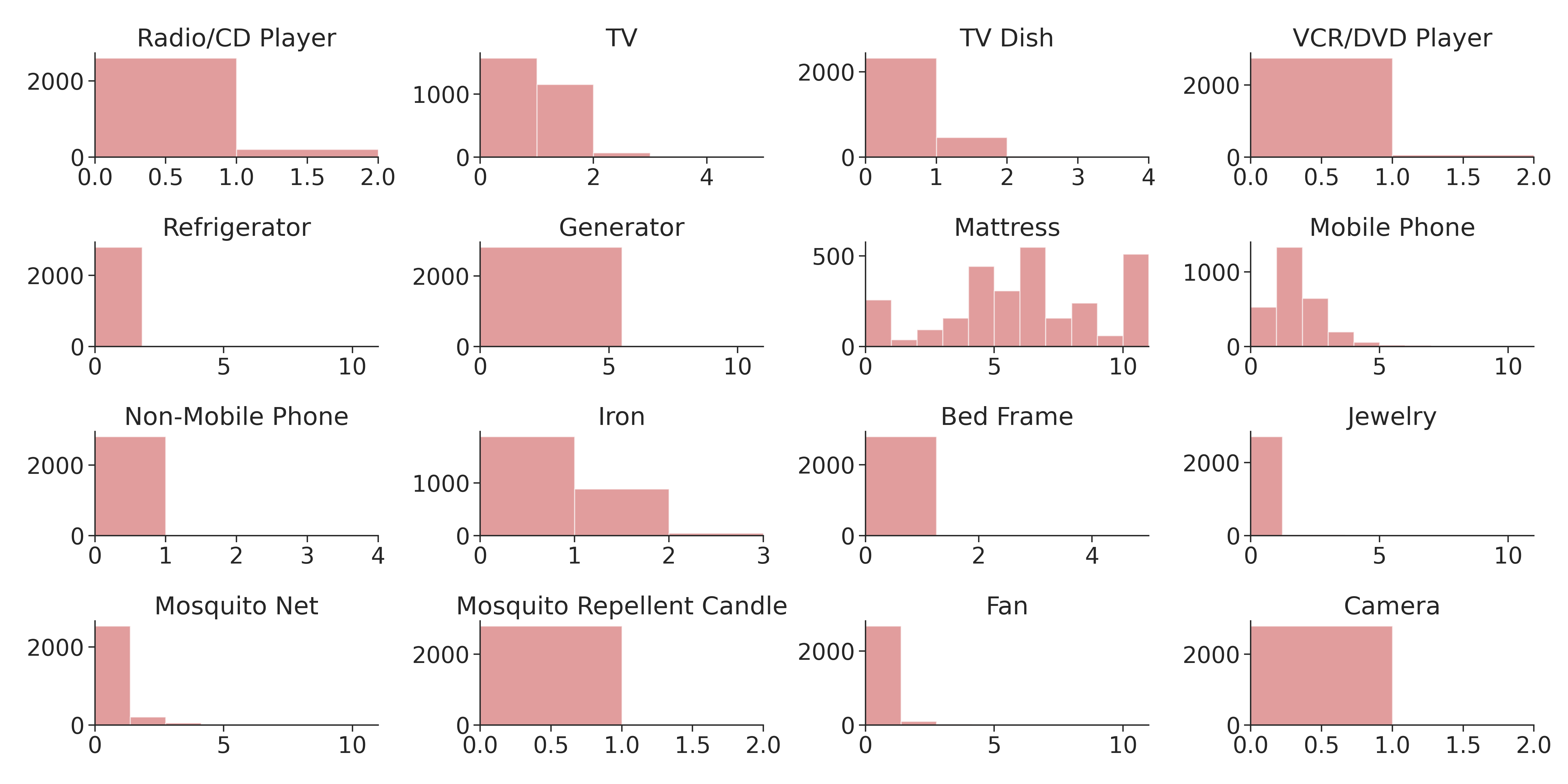}
    \caption{Histograms showing the distribution of each underlying asset used to construct the asset index.}
    \label{appendix:assetindex}
\end{figure}

\begin{figure}[H]
    \centering
    \includegraphics[width=1\textwidth]{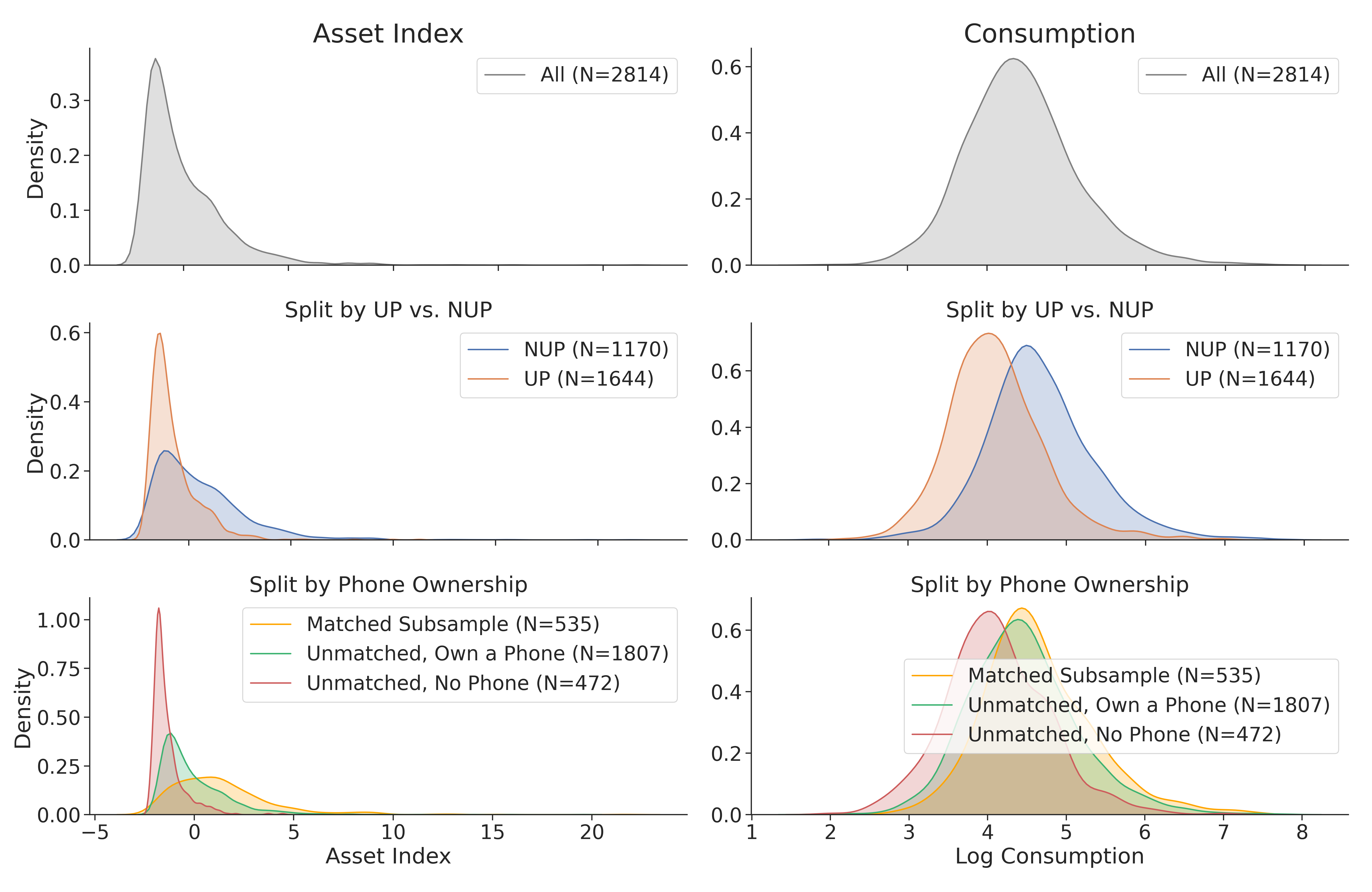}
    \caption{Distributions of asset index and log-transformed consumption, for the entire survey sample, separately for ultra-poor and non-ultra-poor households, and again separately for households in the subsample matched to CDR, households outside of the matched subsample that report owning at least one mobile phone, and households outside of the matched subsample that report not owning a mobile phone.}
    \label{appendix:kdeoutcomes}
\end{figure}

\begin{figure}[H]
    \centering
    \includegraphics[width=1\textwidth]{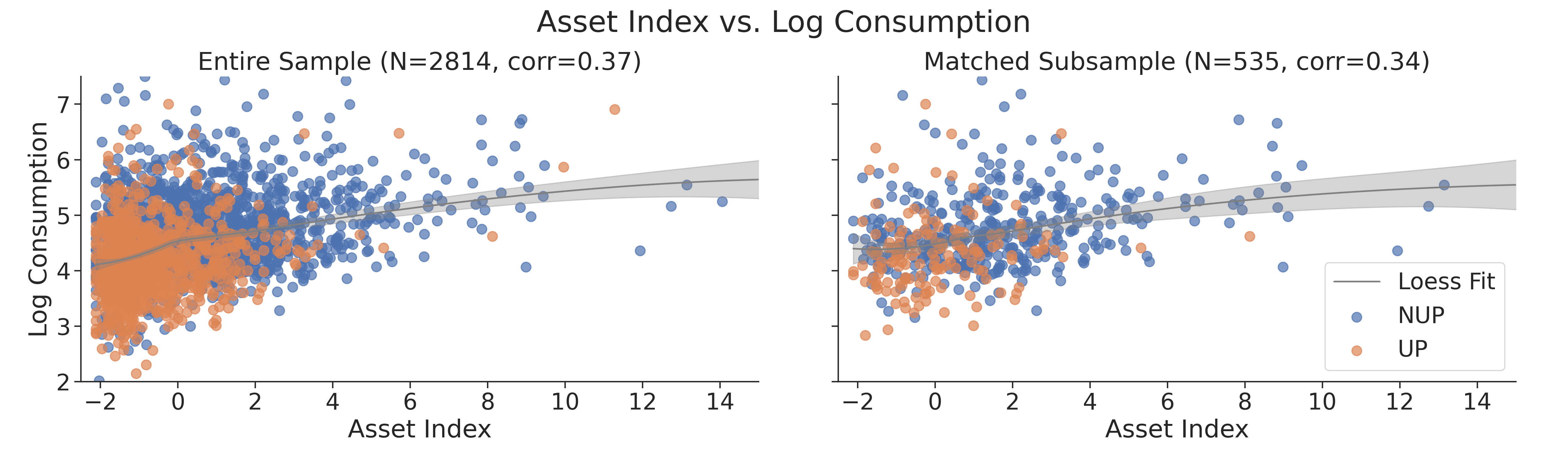}
    \caption{Correlation between asset index and log-transformed consumption, separately for the entire survey sample and the matched subsample. We include the LOESS fit, along with a 95\% confidence interval.}
    \label{appendix:correlations}
\end{figure}

\clearpage
\pagebreak

\begin{table}[H] \centering 
  \caption{Direction of first principal component of asset ownership \label{appendix:firstcomponent}}
\renewcommand{\arraystretch}{1.2}
\begin{threeparttable}
\begin{footnotesize}
\begin{tabular}{lc}
\toprule
                     \textbf{Asset} & \textbf{Magnitude} \\
\midrule
           Radio/CD Player &      0.04 \\
                        TV &      0.37 \\
                   TV Dish &      0.29 \\
            VCR/DVD Player &      0.15 \\
             Refrigerator &      0.25 \\
                 Generator &      0.11 \\
                   Mattress &      0.24 \\
              Mobile Phone &      0.31 \\
          Non-Mobile Phone &      0.06 \\
                      Iron &      0.36 \\
                 Bed Frame &      0.29 \\
                   Jewelry &      0.27 \\
              Mosquito Net &      0.26 \\
 Mosquito Repellent Candle &      0.08 \\
                       Fan &      0.37 \\
                    Camera &      0.16 \\
\bottomrule
\end{tabular} 
\end{footnotesize}
\begin{tablenotes}[normal,flushleft]
\footnotesize
\item \emph{Notes}: The asset index is calculated over the entire 2,814 household sample, without sample weights. We standardize each of the features to zero mean and unit variance before decomposition. The first principal component accounts for 25.28\% of the variation in these standardized features.
\end{tablenotes}
\end{threeparttable}
\end{table} 
\pagebreak

\begin{table}[H] \centering 
  \caption{Feature importances (gradient boosting model) \label{appendix:importances} }
\renewcommand{\arraystretch}{1.2}
\begin{threeparttable}
\begin{footnotesize}
\begin{tabular}{lrlr}
\toprule
                                Feature &  Importance &                              Feature &  Importance \\
\midrule
                      CD WE Call Median &           8 &                IT Recharges Night Min &            3 \\
                     \% At Home WD Night &           7 &                    BOC WD Call Median &            3 \\
             IPC WD Night Call Kurtosis &           7 &                        IT WE Call Min &            2 \\
                     CD Day Call Median &           6 &            BOC WD Night Call Kurtosis &            2 \\
                        IT WE Call Mean &           6 &                 IPC Day Call Kurtosis &            2 \\
                        Churn Rate Mean &           5 &                   \% Nocturnal WD Call &            2 \\
                      IPC Day Call Skew &           5 &                   IT WD Day Text Mean &            2 \\
                 IT Recharges WD Median &           5 &                     CD Night Call Max &            2 \\
                          \% At Home Day &           4 &                       IT WE Call Skew &            2 \\
      \% Initiated Interactions Day Call &           4 &              IPC WE Day Call Kurtosis &            2 \\
   \% Initiated Interactions WD Day Call &           4 &                     IT WE Text Median &            2 \\
                        BOC WD Call Max &           4 &                    \% At Home WE Night &            2 \\
 \% Initiated Interactions WD Night Call &           3 &          Entropy Contacts WE Day Call &            2 \\
                         PPD Night Call &           3 &                    \# Recharges WD Day &            2 \\
              IT Recharges WD Night Min &           3 &                  Entropy Antennas WD  &            2 \\
              IT Recharges Night Median &           3 &                   IPC Night Call Skew &            2 \\
                 IPC WD Night Text Mean &           3 &                 IT WE Night Call Mean &            2 \\
              IT Recharges Day Kurtosis &           3 &                   \# Contacts Day Call &            2 \\
                      IT Night Text Min &           3 &                        CD WD Call Max &            2 \\
                  IT WE Day Text Median &           3 &                      IT Day Call Mean &            2 \\
                         \# Antennas WD  &           3 &                  IT WD Night Text Min &            2 \\
              CD WD Night Call Kurtosis &           3 &                  Entropy Antennas Day &            2 \\
                IPC Night Call Kurtosis &           3 &  \% Initiated Interactions WE Day Call &            2 \\
             IPC WE Night Call Kurtosis &           3 &               CD WE Day Call Kurtosis &            1 \\
                 IPC WE Night Call Skew &           3 &                      IPC Day Call Std &            1 \\
\bottomrule
\end{tabular} 
\end{footnotesize}
\begin{tablenotes}[normal,flushleft]
\footnotesize
\item \emph{Notes}: For our selected machine learning model -- the gradient boosting model used to predict ultra-poor status from CDR features -- we display feature importances for the top 50 features. Feature importances for the gradient boosting model represent the total number of times the feature is used for a split in the entire ensemble of decision trees. We report feature importances when the model is trained on all 535 observations (rather than over cross validation). See Appendix \ref{appendix:abbreviations} for abbreviations in feature names.
\end{tablenotes}
\end{threeparttable}
\end{table} 
\pagebreak

\begin{table}[H] \centering 
  \caption{Details of machine learning models \label{appendix:mldetails} }
\renewcommand{\arraystretch}{1.2}
\begin{threeparttable}
\begin{footnotesize}
\begin{tabular}{l c p{10cm}}
\toprule
                 \textbf{Model} & \textbf{AUC} &                                                                                       \textbf{Top Five Features} \\
\midrule
\\[-.8em]
 Logistic (No Penalty) &      0.53 &                    Reporting \# Records, Active Days, Active Days Day, Active Days Night, Active Days WD \\
 Logistic (L1 Penalty) &      0.66 &                    Reporting \# Records, Active Days, Active Days Day, Active Days Night, Active Days WD \\
         Random Forest &      0.68 &         NOI Out Day Call, NOI Out WD Day Call, Nois  Call, Entropy Contacts Night Call, NOI Out WE Call \\
     Gradient Boosting &      0.68 &  CD WE Call Median, \% At Home WD Night, IPC WD Night Call Kurtosis, CD Day Call Median, IT WE Call Mean \\
\bottomrule
\end{tabular} 
\end{footnotesize}
\begin{tablenotes}[normal,flushleft]
\footnotesize
\item \emph{Notes}: Each row indicates performance (AUC) of a different machine learning algorithm, trained to predict ultra-poor status on the sample of 535 matched households.  AUC is reported as the mean AUC score over 10-fold cross validation. See Appendix \ref{appendix:abbreviations} for details of features.
\end{tablenotes}
\end{threeparttable}
\end{table} 
\begin{table}[H] \centering 
  \caption{Machine learning an asset index \label{appendix:assetbasedpred} }
\renewcommand{\arraystretch}{1.2}
\begin{threeparttable}
\begin{footnotesize}
\begin{tabular}{p{4cm}p{3cm}p{8cm}}
\toprule
                 \textbf{Model} & \textbf{AUC Score} &                                                                                       \textbf{Top Five Features} \\
\midrule
 Logistic (L1 Penalty) &      0.60 &        TV, TV Dish, Fridge, Mattress, Mobile Phone\\
       Random Forest &      0.73 &         Fridge, Iron, Bedframe, Mattress, TV Dish \\
     Gradient Boosting &      0.74 &  Mattress, Bedframe, Fridge, Mobile Phone, TV Dish \\
    
\bottomrule
\end{tabular} 
\end{footnotesize}
\begin{tablenotes}[normal,flushleft]
\footnotesize
\item \emph{Notes}: The asset index benchmark we used is constructed following standard procedures based on principal comnponent analysis (see Table \ref{appendix:assetindex}). However, it is possible that an alternative asset-based predictor, trained using machine learning to predict ultra-poor status directly from the 16 underlying components, could perform better. We test this hypothesis by adapting our machine learning pipeline for identifying the ultra-poor from CDR to the task of identifying the ultra-poor from asset possession.  As with the CDR-based prediction, we evaluate the model over nested cross validation: the model’s predictions are evaluated out-of-sample over 10-fold cross validation, and within each fold hyperparameters are tuned over 5-fold cross validation. We retrain the model on the entire dataset to report hyperparameters and feature importances. Hyperparameters are chosen from the same grid as for the CDR-based models. We display the AUC score and top features for each model.
\end{tablenotes}
\end{threeparttable}
\end{table} 
\pagebreak

\begin{table}[H] \centering 
  \caption{Performance using one, two or three predictor datasets \label{appendix:combined}}
\renewcommand{\arraystretch}{1.2}
\begin{threeparttable}
\begin{tabular}{lc}
\toprule
                     \textbf{Data Sources} & \textbf{AUC} \\
\midrule
           Assets &      0.73 (0.025) \\
                        Consumption &      0.71 (0.000) \\
                   CDR &      0.68 (0.028) \\
            Assets + Consumption &      0.76 (0.017) \\
             Assets + CDR &      0.76 (0.025) \\
                 Consumption + CDR &      0.75 (0.016) \\
                   Assets + Consumption + CDR &      0.78 (0.019) \\

\bottomrule
\end{tabular} 
\begin{tablenotes}[normal,flushleft]
\footnotesize
\item \emph{Notes}: AUC scores for targeting methods using a single data source, pair of data sources, and all three data sources together. Standard deviations are calculated from 1,000 bootstrapped samples of the same size as the original sample, drawn with replacement.
\end{tablenotes}
\end{threeparttable}
\end{table}

\begin{table}[H] \centering 
  \caption{Targeting simulation results for one train-test split\label{onesplit}}
\renewcommand{\arraystretch}{1.2}
\begin{threeparttable}
\begin{footnotesize}
\begin{tabular}{lcccc}
\toprule
 & (1) & (2) & (3) & (4) \\
 \textbf{Targeting Method} &
 \multicolumn{1}{c}{\textbf{AUC}} & 
 \multicolumn{1}{c}{\textbf{Accuracy}} & 
 \multicolumn{1}{c}{\textbf{Precision}} & 
 \multicolumn{1}{c}{\textbf{Recall}} \\
\midrule
                              Random &  0.50  &  0.48 &  0.28 &  0.28 \\
                           Asset Index &  0.68 &  0.63 &  0.33 &  0.33 \\
                      Consumption &  0.74 &  0.74 &  0.53 &  0.53 \\
                                   CDR &  0.75 &  0.70 &  0.47 &  0.47 \\
                              Combined &  0.82 &  0.74 &  0.53 & 0.53 \\

\bottomrule
 \\[-2.5ex] 
\end{tabular} 
\end{footnotesize}
\begin{tablenotes}[normal,flushleft]
\footnotesize
\item \emph{Notes}: Reproduction of main results (Table \ref{resultstable} Panel A) to using a single train-test split (10\% of observations in the training set).
\end{tablenotes}
\end{threeparttable}
\end{table}

\begin{table}[H] \centering 
  \caption{Matching household to multiple phone numbers \label{appendix:individual} }
\renewcommand{\arraystretch}{1.2}
\begin{threeparttable}
\begin{footnotesize}
\begin{tabular}{l c p{10cm}}

\toprule
                 \textbf{Model} & \textbf{AUC} &                                                                                       \textbf{Top Five Features} \\
\midrule

 Logistic (No Penalty) &      0.50 &                    Reporting \# Records, Active Days, Active Days Day, Active Days Night, Active Days WD \\
 Logistic (L1 Penalty) &      0.65 &                    Reporting \# Records, Active Days, Active Days Day, Active Days Night, Active Days WD \\
         Random Forest &      0.67 &         NOI call, NOI Out WE Call, IPC WD Night Call Kurtosis, IPC Night Call Kurtosis, IT Recharges WD Day Min \\
     Gradient Boosting &      0.66 &  Churn Rate Std, CD WE Call Median, IPC WD Night Call Kurtosis, IPC Day Call Skew, \% Initiated Interactions Day Call \\
    
\bottomrule

\end{tabular} 
\end{footnotesize}
\begin{tablenotes}[normal,flushleft]
\footnotesize
\item \emph{Notes}: In our main analysis, for multi-phone households we use only the phone number belonging to the household head (or to a random household member, where no household head is specified), leaving 535 household-level observations. Here we consider instead using machine learning methods to predict individual-level ultra-poverty, with a dataset of 634 individual phone numbers matched to the ground-truth wealth measures for the associated households. We find that the individual-level models are slightly less accurate than the household-level models presented in the main paper, but we focus on the household-level models in the main paper since the household was the unit of targeting in the TUP program.  See Appendix \ref{appendix:abbreviations} for abbreviations in feature names.
\end{tablenotes}
\end{threeparttable}
\end{table}

\begin{table}[H] \centering 
  \caption{Predicting other measures of poverty from CDR \label{appendix:mlresults} }
\renewcommand{\arraystretch}{1.2}
\begin{threeparttable}
\begin{footnotesize}
\begin{tabular}{l c p{10cm}}

\toprule
                 \textbf{Model} & \textbf{$R^2$ or AUC} &   \textbf{Top Five Features} \\
\midrule

\\[-.8em]
\multicolumn{3}{l}{\textit{Panel A: Predicting below poverty line (binary)}}
\\[.3em]

 Logistic (No Penalty) &      0.53 &                    Reporting \# Records, Active Days, Active Days Day, Active Days Night, Active Days WD \\
 Logistic (L1 Penalty) &      0.53 &                    Reporting \# Records, Active Days, Active Days Day, Active Days Night, Active Days WD \\
         Random Forest &      0.56 &  NOI Out Night Call, BOC Night Call Kurtosis, CD Day Call Skew, Nois Night Call, IT Night Call Kurtosis \\
     Gradient Boosting &      0.55 &       IT Night Call Kurtosis, IT  Text Max, Radius Gyration WE Night, Entropy Antennas, NOI Out WD Call \\

\\[-.8em]
\multicolumn{3}{l}{\textit{Panel B: Predicting consumption (continuous)}}
\\[.3em]

 Linear Regression &    -0.21 &  \% Pareto Recharges WE Night, \% Pareto Recharges WE, \% Pareto Recharges Night, Entropy Contacts WD Day Text, PPI WE Night Text \\
  LASSO Regression &    -0.00 &                                                   Reporting \# Records, PPI  Text, PPI Day Text, PPI Night Call, PPI Night Text \\
     Random Forest &    -0.02 &               Churn Rate Mean, IPC WE Night Call Kurtosis, IT Recharges WE Day Skew, IPC WE Night Call Skew, CD WE Call Median \\
 Gradient Boosting &    -0.03 &                  CD WD Night Call Skew, IPC WD Day Text Skew, IT WD Night Call Min, IT WD Night Call Max, IT WE Night Call Max \\

\\[-.8em]
\multicolumn{3}{l}{\textit{Panel C: Predicting asset index (continuous)}}
\\[.3em]

 Linear Regression &    -0.06 &                       IPC  Text Min, IPC WD Text Min, IPC WD Day Text Min, BOC WD Text Min, \% Initiated Conversations WD \\
  LASSO Regression &     0.00 &                                      Active Days WE Day, Active Days WD, Active Days WE, Active Days, Active Days WD Day \\
     Random Forest &     0.00 &              IT Night Call Skew, IPC  Text Min, IT WE Day Call Median, IT WE Call Median, Entropy Contacts WE Night Call \\
 Gradient Boosting &    -0.02 &  IT  Text Median, Entropy Antennas WE, Entropy Antennas WD Night, Entropy Contacts WE Night Call, IT Recharges Night Min \\

\\[-.8em]
\multicolumn{3}{l}{\textit{Panel D: Predicting CWR group (continuous)}}
\\[.3em]

 Linear Regression &     0.01 &                   PPI Night Text, IT Recharges Day Skew, IPC WE Call Min, Active Days WE Night, IT Recharges WD Day Skew \\
  LASSO Regression &     0.05 &                PPI Night Text, Active Days WE Day, Active Days WE Night, IT Recharges WD Day Skew, IT Recharges Day Skew \\
     Random Forest &     0.04 &   \# Contacts WE Day Call, Entropy Contacts WD Night Call, IPC Night Call Kurtosis, \# Contacts WE Call, IT  Call Kurtosis \\
 Gradient Boosting &     0.03 &  IT  Call Kurtosis, IT Recharges Day Skew, \# Contacts WE Day Call, IT Recharges Day Kurtosis, IPC WD Night Call Kurtosis \\
\bottomrule

\end{tabular} 
\end{footnotesize}
\begin{tablenotes}[normal,flushleft]
\footnotesize
\item \emph{Notes}: Machine learning results for predicting: (A) Below-poverty-line status, using consumption data and based on Afghanistan's national poverty line; (B) Total consumption (log-scale); (C) Asset index; and (D) Community Wealth Ranking. Performance is evaluated on the sample of 535 matched households.  Binary metrics (A) are evaluated using the mean AUC score over 10-fold cross validation; Continuous metrics (B-D) are evaluated using the mean $R^2$ score over 10-fold cross validation.  See Appendix~\ref{appendix:abbreviations} for details of features.
\end{tablenotes}
\end{threeparttable}
\end{table}

\begin{table}[H] \centering 
  \caption{Variable costs of different targeting methods \label{appendix:costs}}
\renewcommand{\arraystretch}{1.2}
\begin{threeparttable}
\begin{tabular}{lccc}
\toprule
 & Cost per & Total cost & Fraction of program costs \\ 
Targeting Method & HH screened & of targeting & spent on targeting \\ \midrule
CBT                       & \$2.20                        & \$276,586                     & 2.18\%                                   \\
PMT                       & \$4.00                        & \$502,884                     & 3.97\%                                   \\
Consumption                       & \textgreater{}\$4.00                        & \textgreater{}\$502,884                     & \textgreater{}3.97\%                                   \\
Phone                     & \$0.00                        & \$0                           & 0.00\%                                   \\
\bottomrule
\end{tabular}
\begin{tablenotes}[normal,flushleft]
\footnotesize
\item \emph{Notes}: Costs for the TUP program, based on costs estimated from the literature. The TUP program screened an estimated 125,721 households; benefits valued at \$1,668 were provided to each of the 7,500 beneficiary households for a total benefits distribution of approximately \$12.7 million. The total value of benefits is used to obtain the targeting costs as a percentage of total program costs. For the Phone option, we assume no contact with beneficiaries is required; if contact were required, for instance to collect informed consent, variable costs would increase accordingly.
\end{tablenotes}
\end{threeparttable}
\end{table}

\begin{table}[H] \centering 
  \caption{Costs for CBT and PMT targeting methods obtained from the literature \label{appendix:costliterature}}
\renewcommand{\arraystretch}{1.2}
\begin{threeparttable}
\begin{tabular}{l c c}
\toprule
\textbf{Source}                & \textbf{Location} & \textbf{Cost per household } \\ \midrule
\textit{Panel A: CBT}          &                   &                                      \\
Alatas et al. (2012)           & Indonesia         & \$1.20                               \\
Karlan and Thuysbaert (2019)   & Honduras          & \$1.67                               \\
Karlan and Thuysbaert (2019)   & Peru              & \$1.90                               \\
Schnitzer and Stoeffler (2021) & Burkina Faso      & \$5.60                               \\
Schnitzer and Stoeffler (2021) & Niger             & \$5.40                               \\
Schnitzer and Stoeffler (2021) & Senegal           & \$3.20                               \\
Median                         &                   & \$2.20                               \\ \\
\multicolumn{3}{l}{\textit{Panel B: PMT}}                                                 \\
Alatas et al. (2012)           & Indonesia         & \$2.70                               \\
Karlan and Thuysbaert (2019)   & Honduras          & \$2.62                               \\
Karlan and Thuysbaert (2019)   & Peru              & \$3.05                               \\
Schnitzer and Stoeffler (2021) & Burkina Faso      & \$5.69                               \\
Schnitzer and Stoeffler (2021) & Chad              & \$9.50                               \\
Schnitzer and Stoeffler (2021) & Mali              & \$4.00                               \\
Schnitzer and Stoeffler (2021) & Niger             & \$6.80                               \\
Median                         &                   & \$4.00                               \\ \bottomrule
\end{tabular}
\begin{tablenotes}[normal,flushleft]
\footnotesize
\item \emph{Notes}: Costs per household screened for two targeting methods obtained from three papers in the targeting literature. Costs in \cite{alatas2012} are provided per-village; we use the average of 54 households per village to obtain per-household targeting costs. Cost for the CBT in \cite{karlan2013} is provided as part of the cost for a hybrid CBT and verification approach; although an individual cost for the cBT alone is provided, it is possible this cost excludes some of the mutual costs for the two exercises and is therefore an underestimate of costs of a CBT alone. We use the median of the distribution of targeting costs in our cost analysis. 
\end{tablenotes}
\end{threeparttable}
\end{table}

\pagebreak

\end{document}